\documentclass[12pt]{article}
\usepackage{amsmath,amsfonts,graphicx}

\def\half{\textstyle{\frac{1}{2}}}

\def\H{{\cal H}}

\def\p{\phi}

\def\l{\lambda}

\def\S{\Sigma'}
\def\t{\textstyle}
\def\ZZ{\mathbb Z}

\def\F{{\cal F}}

\def\ra{\rightarrow}
\def\tint{{\textstyle\int}}

\def\s{\hskip.08em}

\def\d{\partial}

\def\b{\begin{eqnarray}}  
\def\e{\end{eqnarray}}    
\def\bn{\begin{eqnarray}}  

\def\<{\langle}
\def\>{\rangle}

\def\no{\nonumber}

\def\k{\kappa}
\def\{{\lbrace}
\def\}{\rbrace}
\title{Taming Nonrenormalizability}
\author{John R. Klauder\footnote{{\sf klauder@phys.ufl.edu}}
\\Department of Physics and Department of Mathematics
\\
 University of
Florida\\
 P.O. Box 118440\\
Gainesville, FL 32611-8440}

\date{ }
\begin{document}
\bibliographystyle{unsrt}
\maketitle

\begin{abstract}
Nonrenormalizable scalar fields, such as $\varphi^4_n$, $n\ge5$, require
infinitely many distinct counter terms when perturbed about the free theory,
and lead to free theories when defined as the continuum limit of a lattice
regularized theory restricted only to arbitrary mass and coupling constant renormalization.
Based on the proposal that functional integrals for interacting nonrenormalizable models do not
reduce to the expression for the free field functional integral as the coupling constant
vanishes -- a proposal supported by the fact that even the set of classical solutions
for such models does not reduce to the set of free field solutions as the coupling constant
vanishes -- it has been conjectured that for nonrenormalizable models the interaction
term acts partially as a hard core eliminating certain fields otherwise allowed by the free
theory. As a consequence, interacting models are continuously connected to a pseudofree theory
that takes into account the hard core as the coupling constant vanishes, and this general view
is supported not only by simple quantum mechanical examples as well as soluble but nonrelativistic
nonrenormalizable models. The present article proposes a pseudofree model for relativistic
nonrenormalizable models about which it is argued that a perturbation expansion of the interaction
is term-by-term divergence free.
\end{abstract}

\section*{Introduction}
Nonrenormalizable quantum field models, such as $\varphi^4_n$, with
a spacetime dimension $n\ge5$, require the introduction of
nonclassical (i.e., $\hbar$-dependent), nontrivial (i.e., other than mass or interaction) counterterms to avoid triviality \cite{AF}. A conventional
regularized perturbation analysis introduces additional counterterms designed to
 cancel divergences as they arise in the
perturbative evaluation of a functional integral about the free theory; but for such theories,
infinitely many distinct counterterms are needed, and therefore this approach is unsatisfactory.
Instead, our procedure chooses the (unconventional) counterterm in order
to cancel the source of the divergences as already identified within the integrand of a functional integral.
Unconventional
counterterms are not out of place since it is highly likely that interacting nonrenormalizable
theories reduce to a pseudofree model different from the free model as the
coupling constant is reduced to zero due to the interaction term partially acting as a hard core. Indeed, consider the Sobolev-like inequality  \cite{book}
   \b \{\tint\p(x)^4\,d^n\!x\}^{1/2}\le C\,\tint\{[\nabla\p(x)]^2+m^2\s\p(x)^2\}\,d^n\!x\;,\label{qq66}\e
valid for $C=4/3$ for $n\le4$, while $C=\infty$ for $n\ge5$ -- which means in the latter case that there
are fields, e.g., $\p_{singular}(x)=|x|^{-p}\,e^{-x^2}$, $n/4\le p<n/2-1$, for which the left side diverges while the
right side is finite. This relation implies that the set of interacting classical solutions does not reduce to
the set of free classical solutions as the coupling constant goes to zero. Even simpler: the
classical action for a single degree of freedom given by
   \b I=\tint\{\s\half[{\dot x}(t)^2-x(t)^2]-\l\s x(t)^{-4}\}\,dt \e
 clearly illustrates the basic principles of a far simpler but analogous hard-core behavior
 and associated pseudofree theory \cite{science}.

We turn to an analysis of the principal subject of this article.
Initially, we choose an $n$-dimensional, periodic, hypercubic,
Euclidean spacetime lattice with
 a lattice spacing $a$, $L$ lattice points on each side, and lattice points
labeled by multi-integers $k=(k_0,k_1,\ldots,k_s)\in {\ZZ}^n$,
where $s=n-1$ is the spatial dimension, and $k_0$ refers to a future
time direction. The lattice-regularized functional integral for the
Schwinger function generating  functional is given by
 \b S(h)\hskip-1.3em&&\equiv M\s\int \exp[\s Z^{-1/2}\s\Sigma_k h_k\s\p_k\s a^n/\hbar
  -I_n(\p,a,N)/\hbar\no\\
  &&\hskip6em-C(\p,a,\hbar)/\hbar]\;\Pi_k\s d\p_k\no\\
  &&\equiv\<\,e^{\t\s Z^{-1/2}\s\Sigma_k h_k\s\p_k\s a^n/\hbar}\;\>\;, \label{t3}\e
  where $\{h_k\}$ is a suitable smooth sequence, and
  the normalization factor $M$ ensures that $S(0)=1$. The
  continuum limit is taken in two steps: (i) The number of
  lattice sites on an edge $L\ra\infty$ and the lattice spacing
  $a\ra0$ so that $La$ remains constant and finite. Thus the
  spacetime volume $V=(La)^n$ as well as the spatial volume (at fixed
  Euclidean time) $V'=(La)^s$ are both finite; (ii) The final step involves $V\ra\infty$ and
  $V'\ra\infty$. In this article we focus on just the first step in the continuum limit and
  assume that both $V$ and $V'$ are sufficiently large. Notationally, we also
  introduce $N=L^n$ and $N'=L^s$, and note that
  sums (and products) such as $\Sigma_k (\Pi_k)$ are over all spacetime,
  while $\Sigma'_k (\Pi'_k)$ are over all space alone at some fixed $k_0$.

  In (\ref{t3}), $Z$ denotes the field strength renormalization factor and
   $I_n(\p,a,N)$ is the naive lattice action,
  \b   I_n(\p,a,N)\equiv\half{\t\sum_k} {\t\sum_{k^*}}\,(\phi_{k^*}-\phi_k)^2\,
 a^{n-2}
 +\half m_0^2{\t\sum_k} \phi_k^2\, a^n
  + \l_0{\t\sum_k}\phi^4_k\,a^n\,,\label{t4}\e
 where $k^*$ denotes any one of the $n$ nearest neighbors to $k$ in
the positive sense, i.e.,
$k^*\in\{\,(k_0+1,k_1,\ldots,k_s)\s,\ldots,\s(k_0,k_1,\dots,k_s+1)\,\}$.
Also in (\ref{t3}) the term
 \b C(\p,a,\hbar)\equiv\half\s\hbar^2\s{\t\sum}_k {\F}_k(\p)\s a^n \e
  represents the
still-to-be-chosen counterterm.

Besides the lattice action, we enlist the help of the associated lattice
Hamiltonian as well as the ground state of that Hamiltonian in our
search for a suitable counterterm and pseudofree theory.
Assuming that the spacetime volume $V<\infty$, it is clear that full
spacetime averages such as $\<\s[{\t\sum}_k \p_k^r\s a^n]^p\s\>$ are
finite, for all positive integers $r$ and $p$, provided that all the
corresponding sharp-time, spatial averages $\<[\s{\t\sum}'_k
\p_k^r\s a^s]^p\>$ are finite; for a proof, see \cite{kla}. In turn,
for a large Euclidean time, the latter expression can be represented as
 \b \<\s[\s{\t\sum}'_k \p_k^r\s a^s]^p\s\>=\int [\s{\t\sum}'_k \p_k^r\s
 a^s\s]^p\,\Psi(\p)^2\,\Pi'_k\s d\p_k\;, \label{ww77}\e
 where $\Psi(\p)$ denotes the ground state of the system.

 Our interest next turns to an analysis of the putative ground state.

\subsection*{Choice of Counterterm}
To understand our basic approach, let us first consider
the idealized example of a free-theory, Gaussian ground-state distribution
 \b \Psi_G(\p)^2\equiv R\s e^{\t-A\Sigma'_k \p_k^2\s a^s}\;, \e
where $A=O(1)$, and, for integral $p\ge0$, let us focus on
the integrals
  \b I_p(A)\equiv R\int [\Sigma'_k\p_k^2\s a^s]^p\s e^{\t-A\Sigma'_k
  \p_k^2\s a^s}\s\Pi'_kd\p_k\;.\e
Such integrals can be evaluated exactly, but we prefer to study them
in an approximate sense by steepest descent methods. To that end we
introduce hyper-spherical coordinates \cite{klau2} defined by
 \b &&\p_k\equiv\kappa\s\eta_k\hskip.2cm,\hskip.3cm\Sigma'_k
  \p_k^2=
 \kappa^2\hskip.2cm,\hskip.3cm
 \Sigma'_k\eta_k^2=1\hskip.2cm,\hskip.3cm\no\\&&\hskip.7cm
 0\le\kappa<\infty\hskip.2cm,\hskip.3cm\
 -1\le\eta_k\le1\;, \e
 and it follows that
   \b I_p(A)=2\s R\int \kappa^{2p}\s a^{sp} e^{\t-A\s\kappa^2\s a^s}\,\kappa^{(N'-1)}d\kappa
    \s\s\delta(1-\Sigma'_k\eta_k^2)\s\Pi'_kd\eta_k\;.\label{ww88} \e
   A steepest descent argument leads to
   \b I_p(A)=O((N'/A)^p)\,I_0(A)\;, \e
   and a perturbation series for $I_1(A)$ about $I_1(1)$ is given by
   \b I_1(A)=I_1(1)-\Delta\s
   I_2(1)+\half\s\Delta^2\s I_3(1)-\cdots\;, \label{w17}\e
   where  $\Delta=A-1$. As $N'\ra\infty$, such a series has
   higher-order, term-by-term
    divergences because the support of the ground-state
   distribution is concentrated on disjoint sets for
distinct $A$ values due, specifically, to the factor
$\kappa^{(N'-1)}$ in the integrand.
   Our goal is to introduce a counterterm that effectively cancels
   the factor $\kappa^{(N'-1)}$, and this can be accomplished, loosely speaking, by
   choosing an idealized  example of a {\it pseudofree model}, about which to expand, with
   a ground-state distribution such that
           \b \Psi_I(\p)^2\propto \kappa^{-(N'-1)}\,e^{\t-A\Sigma'_k\p^2_k\s a^s}\;.\label{w6} \e
           Observe that the use of the distribution $\Psi_I(\p)^2$
           in place of $\Psi_G(\p)^2$ above leads to a series analogous to (\ref{w17})
           that is term-by-term finite.

           Naturally, there are many ways to choose
           a pseudofree ground state that has the desired property expressed in (\ref{w6}),
           and different models will require different versions. In fact, (\ref{w6}) has
           been the starting point to rapidly solve ultralocal models, which are nonrenormalizable quantum field theories without spatial derivatives having a vast symmetry that has  been crucial
           to finding their solution previously \cite{rr}. But ultralocal models are not the subject of this article.

           To deal with relativistic models, we focus on a
      ground state for the pseudofree {\it (pf)} model given by
     \b  \Psi_{pf}(\p)= \label{w7}
     K \s\s\frac{e^{\t-\Sigma'_{k,l}\p_k\s A_{k-l}\s\p_l\s
     a^{2s}/2\hbar
     -W(\p\s\s a^{(s-1)/2}/\hbar^{1/2})/2}}
     {\Pi'_k[\Sigma'_lJ_{k,l}\s\p_l^2]^{(N'-1)/4N'}}\;;\e
     we discuss the constants $A_{k-l}$ and $J_{k,l}$ and the function $W$ below.
This form for the ground state is ensured if we define the
pseudofree theory -- the theory about which a perturbation
   expansion is to take place -- as
      \b &&\hskip-1em S_{pf}(h)=M_{pf}\int \exp[\s Z^{-1/2}\s\Sigma_k h_k\s\p_k\s a^n/\hbar
      -\half{\t\sum_k}
      {\t\sum_{k^*}}
      (\phi_{k^*}-\phi_k)^2\s a^{n-2}/\hbar
      \no\\&&\hskip9em-\half\s\hbar\s{\t\sum_k}{\F}_k(\p)\s a^n]
      \,\Pi_k\s d\p_k\,, \e
      and choose ${\F}_k(\p)$ to yield the denominator in (\ref{w7}). To
      make this connection, we appeal to the associated
      lattice Hamiltonian for the pseudofree model,
\b  &&\H_{pf}= -\half\s{\hbar^2}\, a^{-s}\s{\t\sum_k}'\frac{\t\d^2}{\t\d
\phi_k^2}
  +\half{\t\sum'_k}{\t\sum'_{k^*}}\,(\p_{k^*}-\p_k)^2a^{s-2}\no\\
  &&\hskip3em+\half\s\hbar^2{\t\sum_k}'\F_k(\p)\,a^s -E_0 \,, \e
    and from this association we find that
     \b &&\F_k(\p)
\equiv\frac{1}{4}\s\bigg(\frac{N'-1}{N'}\bigg)^2\s
a^{-2s}\s{\t\sum'_{\s r,\s t}}\s\frac{J_{r,\s k}\s
  J_{t,\s k}\s \p_k^2}{[\S_l\s J_{r,\s l}\s\p^2_l]\s[\S_m\s
  J_{t,\s m}\s\p_m^2]}\no\\
  &&\hskip2cm-\frac{1}{2}\s\bigg(\frac{N'-1}{N'}\bigg)
  \s a^{-2s}\s{\t\sum'_{\s t}}\s\frac{J_{t,\s k}}{[\S_m\s
  J_{t,\s m}\s\p^2_m]} \no\\
  &&\hskip2cm+\bigg(\frac{N'-1}{N'}\bigg)
  \s a^{-2s}\s{\t\sum'_{\s t}}\s\frac{J_{t,\s k}^2\s\p_k^2}{[\S_m\s
  J_{t,\s m}\s\p^2_m]^2}\;. \label{w19}\e
  Irrespective of the choice for $J_{k,l}$,
  we note that: (i) the denominator in the expression for the pseudofree
  ground state specifically leads to the counterterm in the Hamiltonian; (ii) the term in the
  exponent quadratic in $\p$
  is chosen to yield the spatial-gradient term in the
  Hamiltonian (and possibly part of $E_0$), and this requires that $A_{k-l}
  =O(a^{-(s+1)})$; and (iii) the unspecified term $W$ ensures that no
  additional terms (other than the rest of $E_0$) appear in the Hamiltonian. The
  functional form of the argument in $W$ follows from the manner in
  which both $\hbar$ and $a$ appear in the Hamiltonian. In addition,
  note that the quadratic and denominator terms in $\Psi_{pf}(\p)$
  are correct for very large and very small field values, respectively; hence $W$
  is relatively most effective for intermediate field values.

  The choice $J_{k,l}=\delta_{k,l}$ leads to a local covariant
  potential for which $\F_k(\p)\propto 1/\p_k^2$, but it also gives rise to a
  ground-state distribution with incipient normalization divergences at
  $\p_k=0$, for each $k$,
  as $N'\ra\infty$. This behavior is appropriate for an ultralocal model, but not for a relativistic model.
  To overcome that feature, we choose the factors $J_{k,l}$ to provide
  a minimally regularized, lattice-symmetric,
   local spatial averaging in the form
   \b J_{k,\s
l}\equiv\frac{1}{2s+1}\s\delta_{\s k,\s l\in\{k\s\cup \s
k_{nn}\}}\;, \e where $\delta_{k,l}$ is a Kronecker delta. This
notation means that an equal weight of $1/(2s+1)$ is given to the
$2s+1$ points in the set composed of $k$ and its $2s$ nearest
neighbors in the spatial sense only; $J_{k,\s l}=0$ for all other
points in that spatial slice. {\bf [}Specifically, we define
$J_{k,\s l}=1/(2s+1)$ for the points $l=k=(k_0,k_1,k_2,\ldots,k_s)$,
$l=(k_0,k_1\pm1,k_2,\ldots,k_s)$,
$l=(k_0,k_1,k_2\pm1,\s\ldots,k_s)$,\ldots,
$l=(k_0,k_1,k_2,\ldots,k_s\pm1)$.{\bf ]} This definition implies
that $\Sigma'_l\s J_{k,\s l}=1$.

In the continuum limit, it is important to observe that the form of
the counterterm given by (\ref{w19}) leads to a local covariant
potential, albeit an unconventional one.

\section*{The Continuum Limit, and Term-by-term \\Finiteness of a Perturbation Analysis}
      Before focusing on the limit $a\ra0$ and $L\ra\infty$, we
      note several important facts about ground-state averages of the direction
      field variables $\{\eta_k\}$. First, we assume that such averages
      have two important symmetries: (i) averages of an odd number
      of $\eta_k$ variables vanish, i.e.,
      \b \<\eta_{k_1}\cdots\eta_{k_{2p+1}}\>=0\;, \e
      and (ii) such averages are invariant under any spacetime translation, i.e.,
    \b
    \<\eta_{k_1}\cdots\eta_{k_{2p}}\>=\<\eta_{k_1+l}\cdots\eta_{k_{2p}+l}\>\;\e
    for any $l\in{\ZZ}^n$ due to a similar translational
    invariance of the lattice Hamiltonian. Second, we note that
    for any ground-state distribution, it is
    necessary that $\<\s\eta_k^2\s\>=1/N'$
for the simple reason that $\Sigma'_k\s\eta_k^2=1$. Hence,
       $|\<\eta_k\s\eta_l\>|\le1/N'$ as follows from the Schwarz
       inequality. Since $\<\s[\s\Sigma'_k\s\eta_k^2\s]^2\>=1$, it
       follows that $\<\s\eta_k^2\s\eta_l^2\s\>=O(1/N'^{2})$.
       Similar arguments show that for any ground-state distribution
         \b  \<\eta_{k_1}\cdots\eta_{k_{2p}}\>=O(1/N'^{p})\;, \e
         which will be useful in the sequel.

         In discussing the moments below, we remind the reader [see the discussion regarding
          (\ref{ww77})]  that if sharp
         time averages are made finite, then the corresponding spacetime averages
         in the distribution determined by the lattice action will also
         be finite.

\subsubsection*{Field strength renormalization}
         For $\{h_k\}$  a suitable spatial test sequence, we insist
         that expressions such as
         \b \int Z^{-p}\,[\Sigma'_k h_k\s\p_k\,a^s]^{2p}\,\Psi_{pf}(\p)^2\,\Pi'_k\s
         d\p_k \label{w20}\e
         are finite in the continuum limit. Due to the intermediate
         field relevance of the factor $W$ in the pseudofree ground state,
         an approximate evaluation of the integral (\ref{w20}) will
         be adequate for our purposes.
          Thus, we are led to consider
         \b &&\hskip-.3cm K\int Z^{-p}\,[\Sigma'_k
         h_k\s\p_k\,a^s]^{2p}\,\frac{e^{\t-\Sigma'_{k,l}\s\p_k\s A_{k-l}\s\p_l\,
         a^{2s}/\hbar-W}}{\Pi'_k[\s\S_lJ_{k,l}\p_l^2\s]^{(N'-1)/2N'}}\,\Pi'_k\s
         d\p_k\no\\
         &&\hskip.2cm\simeq 2\s K_0\int Z^{-p}\s\k^{2p}\,[\Sigma'_k h_k\s\eta_k\,a^s]^{2p}\\&&\hskip1.4cm\times
         \frac{e^{\t-\k^2\s\Sigma'_{k,l}\s\eta_k\s A_{k-l}\s\eta_l\,a^{2s}/\hbar}}
         {\Pi'_k[\s\S_l J_{k,l}\s\eta^2_l\s]^{(N'-1)/2N'}}\,d\k\,\delta(1-\Sigma'_k\eta_k^2)
         \,\Pi'_k\s d\eta_k\;, \label{f5} \no\e
         where $K_0$ is the normalization factor when $W$ is dropped.
         Our  goal is to use this integral to determine a value for
         the field strength renormalization constant $Z$.
         To estimate this integral we first replace two
         factors with $\eta$ variables by their appropriate
         averages. In particular, the quadratic expression in the exponent
         is estimated by
          \b \k^2\s\Sigma'_{k,l}\s\eta_k\s
          A_{k-l}\s\eta_l\,a^{2s}\simeq\k^2\s\Sigma'_{k,l}\s N'^{\,-1}
          A_{k-l}\,a^{2s}\propto \k^2\s N'\s a^{2s}\s a^{-(s+1)}\;, \e
          and the expression in the integrand is estimated by
          \b [\Sigma'_k h_k\s\eta_k\,a^s]^{2p}\simeq\s
          N'^{\,-p}\,[\Sigma'_k h_k\,a^{s}]^{2p}\;. \e
         The integral over $\k$ is then estimated by first rescaling the variable
         $\k^2\ra\k^2/(N'\s a^{s-1}/\hbar)$, which then leads to an overall integral estimate proportional
          to
         \b  Z^{-p}\,[N'\s a^{s-1}]^{\,-p}\,N'^{-p}\,[\Sigma'_k h_k\,a^{s}]^{2p}\;;\e
         at this point, all factors of $a$ are now outside the
         integral.
         For this result to be meaningful in the continuum limit,
         we are led to choose $Z=N'^{\,-2}\s a^{-(s-1)}$. However, $Z$ must
         be dimensionless, so we introduce a fixed positive quantity
         $q$ with dimensions of an inverse length, which allows us to
         set
          \b Z=N'^{\,-2}\s (q\s a)^{-(s-1)}\;. \e
 \subsubsection*{Mass and coupling constant renormalization}
 A power series expansion of the mass and coupling constant terms in the full spacetime
 distribution leads to two kind of moments given by
  \b \<\s [\s  m_0^2\,\Sigma_k \p_k^2 a^n\s]^p\s\>\;,\hskip3em \<\s [\s  \l_0\,\Sigma_k \p_k^4 a^n\s]^p\s\> \;,\e
   for $p\ge1$, which we treat
  together as part of the larger family governed by
  $\<\s [\s  g_{0,r}\,\Sigma_k \p_k^{2r} a^n\s]^p\s\>$ for integral $r\ge1$.
  We deal with these spacetime moments by considering analogous sharp time moments given by
  \b &&\hskip-.2cm K\int [\s g_{0,r}\Sigma'_k
         \p_k^{2\s r}\,a^s]^{p}\,\frac{e^{\t-\Sigma'_{k,l}\s\p_k\s A_{k-l}\s\p_l\,
         a^{2s}/\hbar-W}}{\Pi'_k[\s\S_lJ_{k,l}\p_l^2\s]^{(N'-1)/2N'}}\,\Pi'_k\s
         d\p_k\no\\
         &&\hskip.7cm\simeq 2\s K_0\int g_{0,r}^p\s\k^{2 r p}\,[\Sigma'_k\s\eta_k^{2 r}\,a^s]^{p}\label{ff5}\\
         &&\hskip1.7cm\times\frac{e^{\t-\k^2\s\Sigma'_{k,l}\s\eta_k\s A_{k-l}\s\eta_l\,a^{2s}/\hbar}}
         {\Pi'_k[\s\S_l J_{k,l}\s\eta^2_l\s]^{(N'-1)/2N'}}\,d\k\,
         \delta(1-\Sigma'_k\eta_k^2)
         \,\Pi'_k\s d\eta_k\;.  \no\e
         The quadratic exponent is again estimated as
         \b \k^2\s\Sigma'_{k,l}\s\eta_k\s
          A_{k-l}\s\eta_l\,a^{2s}\propto \k^2\s N'\s a^{2s}\s a^{-(s+1)}\;, \e
while the integrand factor
         \b [\Sigma'_k\eta_k^{2r}]^p\simeq N'^p\s N'^{-rp}\;. \e
        The same transformation of variables used above precedes the integral
        over $\k$, and the result is an integral, no longer depending on $a$, that is proportional to
          \b g_{0,r}^p \s N'^{-(r-1)p}\s a^{sp}/N'^{rp}\s a^{(s-1)rp}\;.\e
          To have an acceptable continuum limit, it suffices that
              \b g_{0,r}=N'^{(2r-1)}\,(q\s a)^{(s-1)r-s}\,g_r \;,\e
              where $g_r$ may be called the physical coupling factor.
              Moreover, it is noteworthy that
              $Z^r\,g_{0,r}=[N'\s (q\s a)^s]^{-1}\,g_r $,
          for all values of $r$, which for a finite spatial volume $V'=N'\s a^s$ leads to a
    finite nonzero result for $Z^r\s g_{0,r}$. It should not be a surprise that there are no
          divergences for all such interactions because the source of all divergences has been
          neutralized!

    We may specialize the general result established above to the two cases of
    interest to us. Namely, when $r=1$ this last
    relation implies that $m_0^2=N'\s(q\s a)^{-1}\,m^2$, while when
    $r=2$, it follows that $\l_0=N'^{\,3}\s(q\s a)^{s-2}\s\l$. In
    these cases it also follows that $Z\s m_0^2=[\s N'\s (q\s a)^s\s]^{-1}\s m^2$ and
  $Z^2\s \l_0=[\s N'\s (q\s a)^s\s]^{-1}\s \l$,
    which for a finite spatial volume $V'=N'\s a^s$ leads to a
    finite nonzero result for $Z\s m_0^2$ and $Z^2\s \l_0$, respectively.

  \section*{Conclusion}
          For scalar nonrenormalizable quantum field models, we
          have shown that the choice of an unconventional counterterm, but one
          that is proportional to $\hbar^2$ and therefore
          nonclassical,  leads to a formulation for
          which a perturbation analysis  for both the mass term and
          the nonlinear interaction term of the full spacetime averages, expanded about the appropriate pseudofree model, are term-by-term finite. Thanks to the unconventional
          counterterm, it is noteworthy that additive
          renormalization has been everywhere replaced by multiplicative renormalization.
          A natural question to ask is
          how the cancelation of the factor $\k^{(N'-1)}$ in (\ref{ww88}) relates
          to the classical hard core observation regarding fields such as
          $\p_{singular}$ given following (\ref{qq66}); our answer is to observe that
          the existence of the classical hard core tends to suggest there may a hard core in the
          quantum theory as well, but as in all functional integrals over fields, the
          set of fields in the classical domain makes a contribution in the quantum theory
          of measure zero.

          Alternative insight into such models may possibly be
          obtained by Monte Carlo studies of the full,
          nonperturbative model including the special counterterm; for
          a preliminary discussion of such an approach, see \cite{kl}.

  \end{document}